\documentclass{aa}

\def\var{OGLE-BLG-ECL-157529}

\usepackage{graphicx}
\usepackage{tabularx}
\usepackage{txfonts}
\begin{document}

  \title{Model for the long and orbital brightness variability of the $\beta$ Lyrae type binary OGLE-BLG-ECL-157529}
   \subtitle{}

   \author{R.E.\,Mennickent
          \inst{1}
                   \and
          G.\,Djura\v{s}evi\'c
          \inst{2,3}
          }

   \institute{Universidad de Concepci\'on, Departamento de Astronom\'{\i}a, Casilla 160-C, Concepci\'on, Chile\\
              \email{rmennick@udec.cl}
         \and
             Astronomical Observatory, Volgina 7, 11060 Belgrade 38, Serbia
           \email{gdjurasevic@aob.rs}
                     \and
                     Issac Newton institute of Chile, Yugoslavia Branch, 11060, Belgrade, Serbia       
             }
   \date{Received XX XX, 2021; accepted XX XX, 2021}

 \titlerunning{Disk model and evolution in a $\beta$ Lyrae-type binary }
\authorrunning{Mennickent \& Djura\v{s}evi\'c}

  \abstract
   {Some  close binaries of the $\beta$ Lyrae type show photometric cycles longer than the orbital one, which are possibly related to changes in their accretion disks.}
   {We aim to understand the short- and long-scale morphologic changes observed in the light curve of the eclipsing system OGLE-BLG-ECL-157529. In particular, we want to shed light on the contribution of the disk variability  to these changes, especially those related to the long cycle, occurring on timescales of hundreds of days. }
   {We studied $I$-band
Optical Gravitational Lensing Experiment (OGLE) photometric times series spanning 18.5 years, constructing disk models by analyzing the orbital light curve at 52 different consecutive epochs.  An optimized  simplex algorithm was used to solve the  inverse problem by adjusting the light curve with the best stellar-orbital-disk parameters for the system. We applied an analysis of principal components to the parameters to evaluate their dependence and variability.  We constructed a description of the mass transfer rate in terms of disk parameters.}
   {We find that the overall light variability can be understood in terms of a variable mass transfer rate and variable accretion disk. 
   The system brightness at orbital phase 0.25 follows the long cycle and is correlated with the mass transfer rate and the disk thickness. The long-cycle brightness variations can be understood in terms of differential occultation of the hotter star by a disk of variable thickness. Our model fits the overall light curve during 18.5 years well, including epochs of reversal of main and secondary eclipse depths. The disk radius cyclically change around the tidal radius, decoupled from changes in the mass transfer rate or system brightness, suggesting that viscous delay might explain the non-immediate response. Although the disk is large and fills a large fraction of the hot star Roche lobe, Lindblad resonance regions are far beyond the disk, excluding viscous dissipation as a major source of photometric variability.     }
   {}


   \maketitle
%

\section{Introduction}

\begin{figure*}
\scalebox{1}[1]{\includegraphics[angle=0,width=18cm]{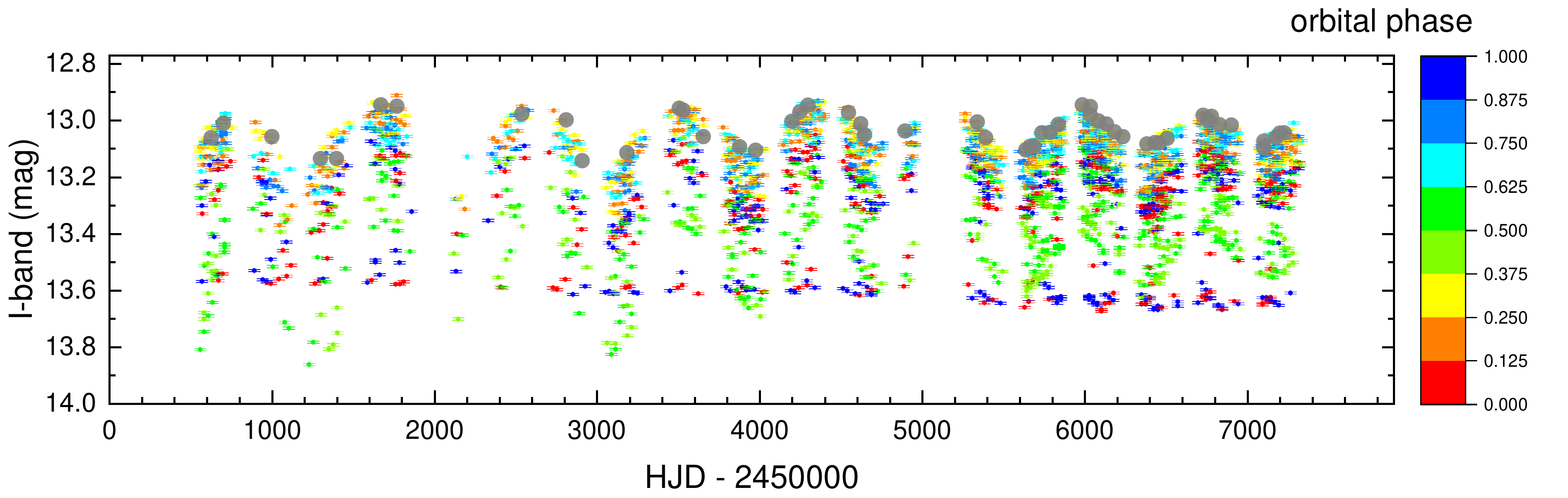}}
\caption{OGLE $I$-band light curve of OGLE-BLG-ECL-157529. Colors indicate different data ranges. Gray dots show the magnitude at orbital phase 0.25 for the 52 consecutive data subsamples.  }
\label{fig:lightcurve}
\end{figure*}

Algol-type binaries are characterized by one of the stellar components having inverted its original mass ratio 
following a Roche lobe overflow and, over time, showing a less massive star that is more evolved than the more massive component. Defined in this way, Algol binaries indicate a particular stage  of binary star evolution. On the other hand, Algol-type eclipsing binaries (also known as  EA-type)  refer to the morphology of the light curve, those showing a flat plateau between eclipse minima, contrary to the rounded inter-eclipse section of $\beta$ Lyrae type eclipsing binaries (also referred as EB-type).  In this paper, we use the term Algols to refer to binaries that have experienced mass ratio reversals in the past, making up a sample that includes the well-studied system $\beta$ Lyrae.
 
In many Algols, an accretion disk is observed surrounding the more massive (gainer) component. The disk  is fed by a gas stream falling ballistically from the inner Lagrangian point  towards the gainer. A "hot spot"  is eventually formed in the stream-disk impact region.  The first direct evidence of the disk was found in $\beta$ Lyrae interferometric images of the Center for High Angular Resolution Astronomy (CHARA)  \citep{2008ApJ...684L..95Z}, while indirect imaging methods such as Doppler tomography have revealed the disk and their gaseous structures in some systems  \citep{2014ApJ...795..160R}. The nature of these disks is not well understood and still is the subject of ongoing studies. Accretion  towards  a near-main sequence star is not so efficient as  towards a compact object, as tends to happen in cataclysmic variables. This means that the luminosity observed in some well-developed disks of Algols is not driven by accretion but probably via heating from photons emitted by the hot central star \citep{2016MNRAS.461.1674M, 2016A&A...592A.151V}.

The fact that the disk is probably a kind of accretion-deccretion (AD) disk was pointed out by \citet{2018ApJ...869...19W}, who proposed a model for $\beta$ Lyrae including a self-gravitating, semi-transparent analytic disk model that considers the critical rotation for the more massive star.  While the outer parts of the disk are subject to the mass inflow from the stream falling from the inner Lagrangian point, the central parts might be subject to inward-outward transfer of angular momentum from the rapidly spinning star. 

The condition of critical rotation might be satisfied by the mass gainer stars of double periodic variables (DPVs) \citep{Mennickent2003, Mennickent2017}. These close binaries are mostly Algols with B-type main sequence gainers whose radii are large enough to allow for a near-tangential impact of the gas stream favoring the acceleration to critical rotation  \citep{Mennickent2016}. The peculiar and puzzling condition of DPVs is the presence of a as yet unexplained long super-orbital photometric periodicity of typical $I$-band amplitude 0.1-0.2 mag. More than 200 DPVs have been reported in the Galaxy and the Magellanic Clouds \citep{Mennickent2003, 2010AcA....60..179P, Pawlak2013, Mennickent2016, Rojas} and a magnetic dynamo operating in the donor star has been proposed as the mechanism modulating mass transfer and, as a result, the system brightness \citep{2017A&A...602A.109S}.  Evidence for relatively massive AD disks could be present in DPVs since most of them show light curves similar to $\beta$ Lyrae. The presence of rapidly rotating B-type components might favor the concentration of mass in the disk around the gainer. It is well known that single B-type stars are the fastest rotators among main sequence stars and this can be understood in terms of internal angular-momentum re-distribution during nuclear evolution \citep{2014A&A...570A..18G}. An open question remains regarding whether this structural characteristic operates in B-type gainers that are subject to rapid acceleration by mass transfer.

From the evolutionary point of view, disks are expected during episodes of mass transfer in some close binaries containing B-type components \citep{2016A&A...592A.151V} and  mass is expected to escape from the system due to the critical rotation of the gainer star at some epochs \citep{2013A&A...557A..40D}. However, evidence for such a mass loss is scarce \citep{2015A&A...577A..55D}. In $\beta$ Lyrae, where a long cycle of 282 days is observed,  a bipolar wind of unknown origin has been revealed by interferometry and optical spectroscopy \citep{1996A&A...312..879H},  whose properties have been quantified by 
 \citet{2018A&A...618A.112M} and 
\citet{2021A&A...645A..51B}. V\,393 Scorpii is another DPV where a bipolar wind has been proposed to explain  spectroscopic observations of high resolution and high signal-to-noise \citep{2012MNRAS.427..607M}. Interestingly, few observational constraints for the long-term variability of $\beta$ Lyrae type binaries have been provided, however, dedicated photometric surveys provide unique opportunities of exploring this variability \citep{Garces2018, Pawlak2013, 2010AcA....60..179P}.

In this paper, we present a study of the eclipsing binary DPV \var\ ($\alpha_{2000}$=17:53:08.33, $\delta_{2000}$=$-$32:46:27.0, $I$= 13.035 mag, $V$= 14.829 mag, \citet{Soszynski2016}). This object shows a $\beta$ Lyrae type orbital light curve, consists of a pair B6 dwarf $+$ K2 giant and is located at a distance of 3024 $\pm$ 406 pc in the direction of the Galactic Bulge. 
It is characterized by  the orbital period of 24\fd80091 $\pm$  0\fd00044 and a  long cycle decreasing in length, from around 900 days to around 800 days, and amplitude, over 18.5 years of observations \citep{2020A&A...641A..91M}. These authors  found that 
the overall light curves can be understood in terms of a variable accretion disk and that the disk is larger and thicker at long cycle minimum and this effect is more pronounced when the long cycle has a larger amplitude and longer cycle length.  The authors constructed a model for the system indicating changes in the temperatures of the hot spot and the bright spot of the disk during the long cycle, and also in the position of the bright spot; hence,  the system represents a unique opportunity for studying the evolution of an accretion disk in a  double periodic variable
and to constrain future models for AD disks in close binaries of the Algol type with $\beta$ Lyrae type characteristics. 

The above authors modeled the system on two epochs, the minimum and maximum of the long cycle, since these two epochs presented an inversion of the relative depths of the two eclipses.They 
 did not study the whole physical content of the large dataset available.  In this paper, we study  the disk evolution of this system further by modeling the light curve in many subsequent epochs to build the overall picture of disk and system evolution during the whole time baseline  span of almost 19 years. Following this approach, we are able to gain new insights on the nature of the long cycle and disk evolution.

\section{Photometric data and light curve}

The photometric time series analyzed in this study has already been described by  \citet{2020A&A...641A..91M}. It consists of 2606 $I$-band data points  taken from The 
Optical Gravitational Lensing Experiment (OGLE, Fig. \ref{fig:lightcurve}): 
OGLE-II  \citep{2005AcA....55...43S}
and OGLE-III/IV.
The OGLE-IV project is described by \citet{2015AcA....65....1U}, while the analysis of photometric uncertainties in the OGLE-IV Galactic Bulge database is given  by \citet{2016AcA....66....1S}. 
The whole dataset, summarized in Table \ref{tab:observinglog}, spans a time interval of  6781 d or 18.5 yr.
In this paper, we use the ephemeris for the occultation of the hotter star   provided by \citet{2020A&A...641A..91M}:
\small
\begin{eqnarray}
\rm HJD&=&\rm (245\,6695\fd918 \pm \rm 0\fd059) + \rm (24\fd80091 \pm \rm 0\fd00044)\, E.
\end{eqnarray}
\normalsize
The light curve variability cannot be understood in terms of a  binary system alone, and the presence of circumstellar matter around the gainer is suggested 
from the changing depth of the minima around phase 0.5 (Fig. \ref{fig:lightcurve}). The cooler star (the donor) is eclipsed by a variable physical structure. However, the depth of eclipse at phase 0.0 is rather 
stable, smoothly increasing during the whole time baseline. These observational facts were interpreted as having been attributed to a  circumstellar disk  that was larger at the beginning of the time series and smaller at the end,  and which surrounds the hotter star \citep{2020A&A...641A..91M}. This scenario can explain even the epochs  when the eclipse at phase 0.0 turns to be dimmer than in phase 0.5 as 
 a result of a larger occultation on the part of the gainer by a thicker disk.

\begin{table}
\centering
\caption{Summary of photometric observations.  }
\label{tab:observinglog}
\begin{tabular}{lcccc} 
\hline
\small
band/Data-Base &  N      & $HJD'_{start}$      &  $HJD'_{end}$   & Mag      \\
\hline
I / OGLE-II    & 346     &   551.77073   &  1858.52147  &   13.213 \\
I / OGLE-III     & 795     &   2117.76494  &  4955.73490  &   13.221  \\
I / OGLE-IV      & 1465    &   5261.84891  &  7332.50545  &   13.213  \\
\hline
\vspace{0.05cm}
\end{tabular}
Note: The number of measurements, starting and ending times, and average magnitude are given. 
HJD' = HJD-2450000. The uncertainty of a single measurement varies between  4 and 6 mmag.
\end{table}

\section{Methodology and light curve analysis}

In order to study the photometric changes, we divided the whole dataset in subsamples of 52 consecutive data points.  This number was chosen because it provides clean orbital light curves with not much interference of the long cycle. The selection of a different number of consecutive segments does not change the results of this paper, keeping in mind that it must be large enough to include data to  adequately cover the orbital light curve and short enough to exclude the perturbing influence of the long cycle.

\subsection{The light curve model}

An optimized simplex algorithm was used to solve the inverse problem adjusting the light curve with the best stellar-orbital-disk parameters for the system. The  basic  elements  of  the  model, together  with  the  light-curve synthesis procedure, 
can be found in the literature \citep{1992Ap&SS.196..267D, 1996Ap&SS.240..317D}, along with new and improved versions \citep{2008AJ....136..767D}, which have been applied to several close binaries in the past \citep[e.g.,][]{2013MNRAS.432..799M, 2018MNRAS.476.3039R, 2020A&A...642A.211M}.

The model describes the total flux of the binary as the sum of the stellar fluxes and the radiation emerging from an optically thick accretion disk surrounding the hotter star and it includes the geometrical effects produced by the observer inclination. The disk light contribution is calculated in terms of local Planck radiation functions at given temperatures and does not consider the  calculation  of  the  radiative  transfer. However, the simple reflection effect as well as the limb and gravity darkening are included.  Details of the model are described in an earlier publication \citep{2008AJ....136..767D} and in the references given there.

 We use the stellar and system parameters derived by \citet{2020A&A...641A..91M}, which  allow us to vary the parameters of the disk (see Table \ref{tab:system}),. 
 The disk is characterized by its radius $R_d$, a vertical thickness at the central and outer edge ($d_c$ and $d_e$, respectively), and a radial-dependent temperature profile:  
\begin{equation}
T(r) = T_{d} \left(\frac{R_{d}}{r}\right)^{a_{T}}, 
\end{equation}

where $T_{d}$ is the disk temperature at its outer edge ($r=R_{d}$) and $a_{T}$ is the temperature exponent ($a_{T} \leq 0.75$). The value of exponent $a_{T}$ shows how close is the radial temperature profile to the steady-state configuration ($a_{T}=0.75$). The radial dependencies of the accretion disk temperature show that the surface of the disk is hotter in the inner regions, and that disk is cooled as we move away from the center. The above power-law differs from the one used by \citet{2020A&A...641A..91M} for the first study of this system.  It appears that a power law is a better description for modeling the $\beta$ Lyrae disk  \citep{2018A&A...618A.112M}.

In addition, the model considers two  shock regions in the disk edge: one is the hot spot, located around the hypothetical impact point between the gas stream ejected from the inner Lagrangian point and the disk, while the other one is a bright spot,  located somewhere in the disk edge.  Both regions represent regions in the disk with variable thickness and temperatures with respect to the surrounding disk and they have been detected in Doppler maps of Algols and also in hydrodynamical simulations of gas flows in close binary stars   \citep[e.g.,][]{1996ApJ...459L..99A, 2000A&A...353.1009B, 2012ApJ...760..134A}. Both spots were found after the analysis of the orbital light curve of $\beta$ Lyrae \citep{2013MNRAS.432..799M}. Interferometric and polarimetric studies seem to confirm the existence of the hot spot in this system \citep{2012ApJ...750...59L, 2018A&A...618A.112M}. While the origin of the hot spot is rather clear, the origin of the bright spot may be attributed to vertical oscillations of the gas at the outer edge of the accretion disk due to interactions with the stream of matter \citep{2017ARep...61..639K}. 
 We notice that our model with a single hot spot on the edge of the disk cannot successfully describe the relatively complex shape of the constantly variable light curves as it requires the addition of the bright spot.

In our model, both regions are characterized by their relative temperatures, $A_{hs} \equiv T_{hs}/T_d$ and $A_{bs} \equiv T_{bs}/T_d$, for their angular dimensions, $\theta_{hs}$ and $\theta_{bs}$ and their angular locations measured from the line joining the stars in the direction of the orbital motion, $\lambda_{hs}$ and   $\lambda_{bs}$. Finally, $\theta_{rad}$ is the angle between the line perpendicular to the local disk edge surface and the direction of the hot spot maximum radiation.
We also consider, as a measure of the disk radius, the parameter $F_d$ $\equiv$ $R_d$/$R_{yk}$, where  $R_{yk}$ is defined by \citet{1992Ap&SS.196..267D} as the distance between the center of the hot star and its Roche lobe in the direction perpendicular to the line joining the star centers.  Defined in this way, the disk is only stable within the limit $F_d$ $\leq$ 1, while with a larger disk radius the material outside the Roche oval would go to the surrounding space and, most likely, in the area of the Lagrange equilibrium point, L$_{3}$, escaping from the system and forming a kind of circumbinary envelope.
The gravity-darkening coefficients of the  stars are fixed to ${\rm \beta_1= 0.25}$ and ${\rm \beta_2= 0.08}$  and the albedo coefficients are ${\rm A_h= 1.0}$ and ${\rm A_2= 0.5}$.

 It is worth noticing that while past models for $\beta$ Lyrae assumed a fixed geometry \citep{2018A&A...618A.112M, 2021A&A...645A..51B},
 our model allows for a variable geometry of the accretion disk, being, in principle, superior with regard to studying the system variability. On the other hand, the treatment
of radiative transfer in the circumstellar medium is more complex in the aforementioned models, including synthetic
spectra for stellar components, local thermodynamic equilibrium, opacities, and spectral lines, while even some of them include the analysis of interferometric data allowing us to test the presence of outflows and jets. Our model does not deal with the treatment of radiative transfer and works with single-band photometric time series and, thus, it is limited in this sense.

The result of the light curve modeling is shown in Table \ref{tab:data}. Some of the single-data-segment fits are shown in Fig. \ref{fig:fits}, illustrating the  uncertainties of the measurements, typical orbital variability, long-term variability, and fit quality.  We found some observed minus calculated residuals  that are larger than individual datapoint errors, which may be explained as underestimated formal errors or sources of variability not 
considered by our model. Another source of the discrepancy could be the fact that individual light curves span several orbital cycles; thus, part of the long-term variability is introduced here.

\begin{table}
\centering
\caption{Certain stellar, orbital and disk parameters, including orbital separation, $a_{orb}$, and system inclination, $i$.  }
\label{tab:system}
\begin{tabular}{lclc} 
\hline
 $M_1$   & 4.83  $\pm$ 0.3 & $T_2$                      & 4400  (fixed)                          \\
 $M_2$   & 1.06  $\pm$ 0.2 & log \ $g_1$             & 3.82  $\pm$ 0.1 \\
 $R_1$   & 4.48  $\pm$ 0.2 &  log \ $g_2$             & 2.02  $\pm$ 0.1 \\
 $R_2$   & 16.6  $\pm$ 0.2 &  $a_{orb}$    & 64.6 $\pm$ 0.3\\
 $T_1$   &    14000 $\pm$ 500                       &       $i$             & 85.5  $\pm$ 0.3 \\
 $P_o$  &24\fd80091 $\pm$  0\fd00044     &$R_d$     &(27.4 $\rightarrow$  33.2) $\pm$ 0.3 \\
 \hline
 \vspace{0.05cm}
\end{tabular}
Note: Indexes 1 and 2 refer to hot and cool stellar components. 
Inclination is given in degree,  temperatures in Kelvin and masses and lengths in M$_{\odot}$ and R$_{\odot}$, respectively. 
The data are from  \citet{2020A&A...641A..91M}.
\end{table}
\normalsize

\subsection{A simple model for the mass transfer rate}

We also calculated an approximation for the mass transfer rate, $\dot{M}$, using the following assumptions. In a semidetached binary ongoing Roche lobe overflow with mass transfer rate, 
$\dot{M_2}$,  towards a star of mass $M_1$,  the luminosity of the disk hot spot is (following \citet{1995CAS....28.....W}, Eq. 2.21a):

\begin{eqnarray}
L_{hs} = \eta \tau {\rm G} \frac{{M_{\rm_1}} \dot{M_{\rm_2}}}  {{R}_{\rm d}},
\end{eqnarray} 

\noindent
where $R_d$ is the disk radius. The $\eta$ factor (between 0 and 1) takes into  account the fact that  since the stream impacts obliquely, only the velocity component 
perpendicular to the disk outer edge participates in the transformation to kinetic energy. In addition, $\tau$ is another factor that considers 
that the fall of material  towards the disk is not from infinity, but from the inner Lagrangian point, and  G is the gravitational constant. 

On the other hand, in considering the hot spot radiating as a blackbody, its luminosity can be expressed as:

\begin{eqnarray}
 L_{hs} \approx  \sigma  T_{hs}^{4} d_e l = \sigma [A_{hs}T_{d}]^{4} d_e l ,
\end{eqnarray} 

\noindent
where $l$ is a measure of the arc subtended by the hot spot along the outer disk edge:

\begin{eqnarray}
l = {{R}_{d}} \theta_{hs},
\end{eqnarray} 

\noindent
We note that in the absence of any measurement of the radial extension of the hot spot, we have used its extension along the disk border.
Hence, during two stages labeled as $i$ (initial) and $f$ (final), and assuming that $\eta$ and $\tau$ are constant, the mass transfer rate satisfies the following:

\begin{eqnarray}
 \frac{\dot{M}_{c,f}}{\dot{M}_{c,i}}= \frac{  {{R}^2_{d,f}} [A_{hs,f}T_{d,f}]^{4} d_{e,f}  \theta_{hs,f} }{{{R}^2_{d,i}} [A_{hs,i}T_{d,i}]^{4} d_{e,i} \theta_{hs,i}}.
\end{eqnarray} 

This equation allows us to estimate relative mass transfer rates at different epochs for a given system. In order to normalize our determined values of $\dot{M}$, we used as initial state the one with the minimum radius, corresponding to the data string  LC = 49, with a mean HJD' = 7098.91. These values are also shown in Table \ref{tab:data},   and an uncertainty of about 50\% is derived from the classical formula of error propagation. 
The above line of reasoning also allows us to compare different systems, $A, B,$ with their disks. To obtain $\dot{M_A}/\dot{M_B}$  we need to change indexes $f, i$ by labels $A, B$ and multiply Eq.\,6 by a factor of $M_B/M_A$.

\subsection{Principal component analysis }

Once we derived the set of disk parameters for the light curves, we ran a principal component analysis (PCA) on the whole dataset. This tool allows us to 
reduce the dimensionality of a dataset, extracting the more significant variables and exploring the correlations between them. The PCA has been extensively used in astronomy 
as a multivariate analysis tool starting from the work on classification of stellar spectra \citep{1964MNRAS.127..493D}.  The technique has recently been  used,  for example,  in the analysis and classification of variable stars \citep{2009A&A...507.1729D} and dissecting the molecular structure of the Orion B cloud \citep{2017A&A...599A.100G}. We used the PCA tool implemented in the OriginLab\footnote{OriginPro, Version 2021. OriginLab Corporation, Northampton, MA, USA.} software.

\section{Results}

\begin{figure}
\scalebox{1}[1]{\includegraphics[angle=0,width=8.5cm]{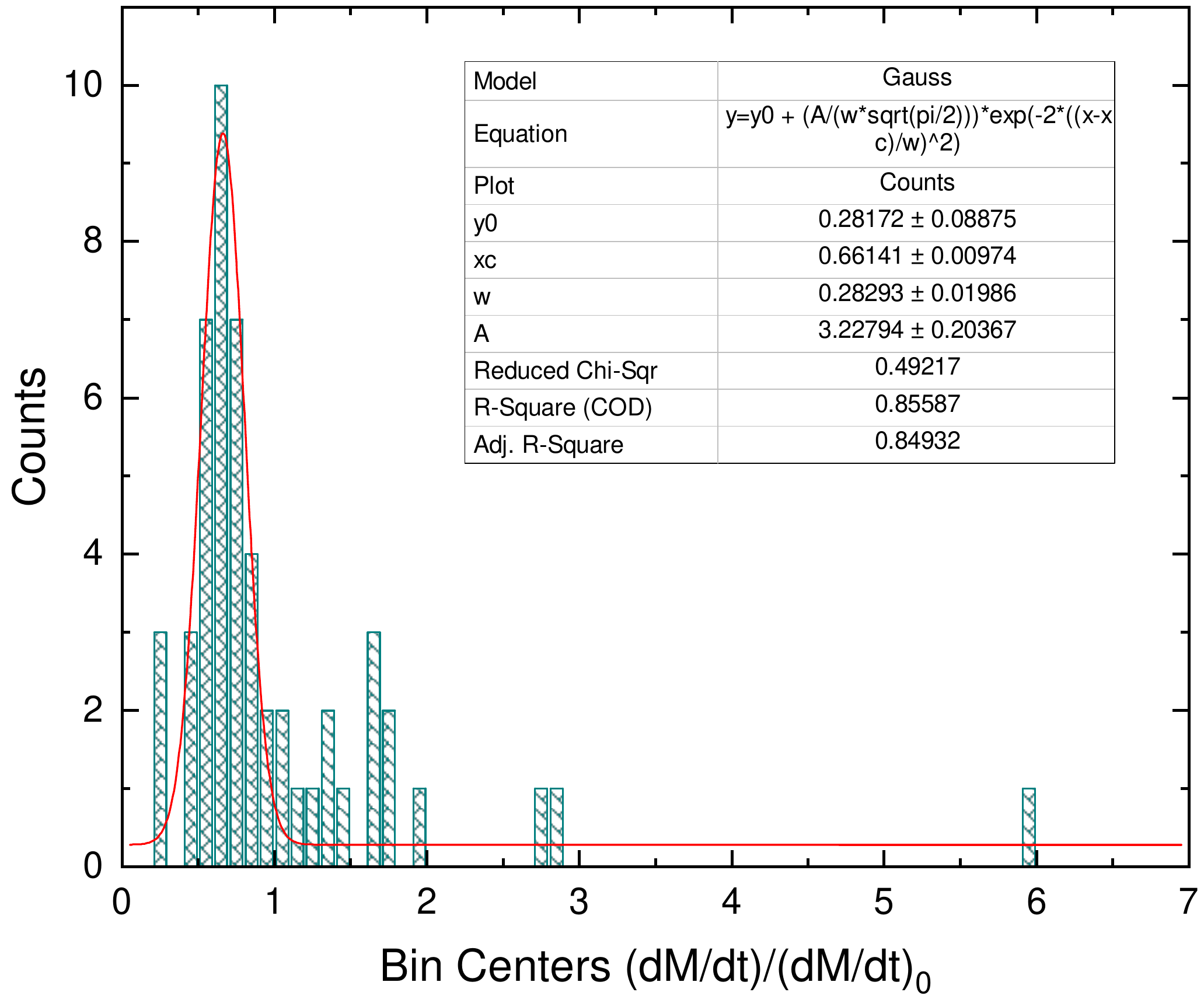}}
\caption{  Histogram of the calculated $\dot{M}/\dot{M}_0$ listed in Table  \ref{tab:data}.}
\label{fig:dmdtNV2}
\end{figure}

  We find a range of normalized $\dot{M}$ values [0.22:5.98] with mean 1.02 and standard deviation 0.90 (Fig. \ref{fig:dmdtNV2}). Changes by a factor 27 between minimum and maximum are observed although this range is based only in a single light curve dataset. The  normalized $\dot{M}$ distribution can be fit by a Gaussian function centered at 0.66 $\pm$ 0.01, area 3.23 $\pm$ 0.20, and full width at half maximum $FWHM\, (\equiv w \sqrt{In\,4}$) = 0.33 $\pm$ 0.02, where $w$ is a parameter defining the width of the gaussian according to the definition given in Fig. \ref{fig:dmdtNV2}. Some  
outliers are found suggesting epochs of large mass transfer rate. In order to examine this possibility, we identified the datasets with the largest  $\dot{M}$  values, namely, in decreasing order of   $\dot{M}$, LC 5, 4, and 11.
These light curves are relatively well sampled compared with other light curves, hence the large $\dot{M}$ values cannot be due to poor sampling of the orbital cycle.  On the contrary, we identified these epochs with relatively bright disks, whose fluxes are larger than the gainer fluxes  (Fig. \ref{fig:fignew}).

\begin{figure*}
\scalebox{1}[1]{\includegraphics[angle=0,width=8.5cm]{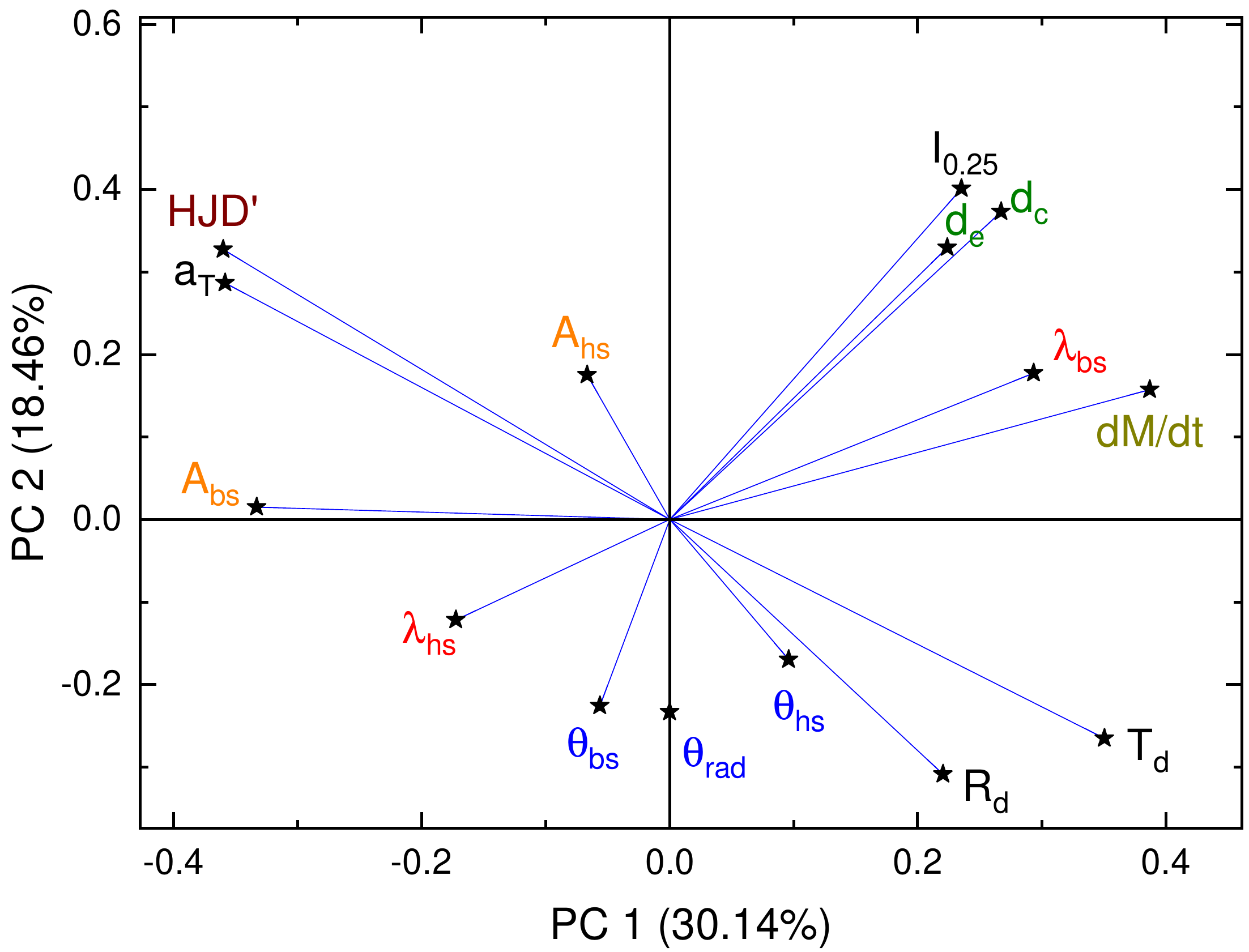}}
\scalebox{1}[1]{\includegraphics[angle=0,width=9cm]{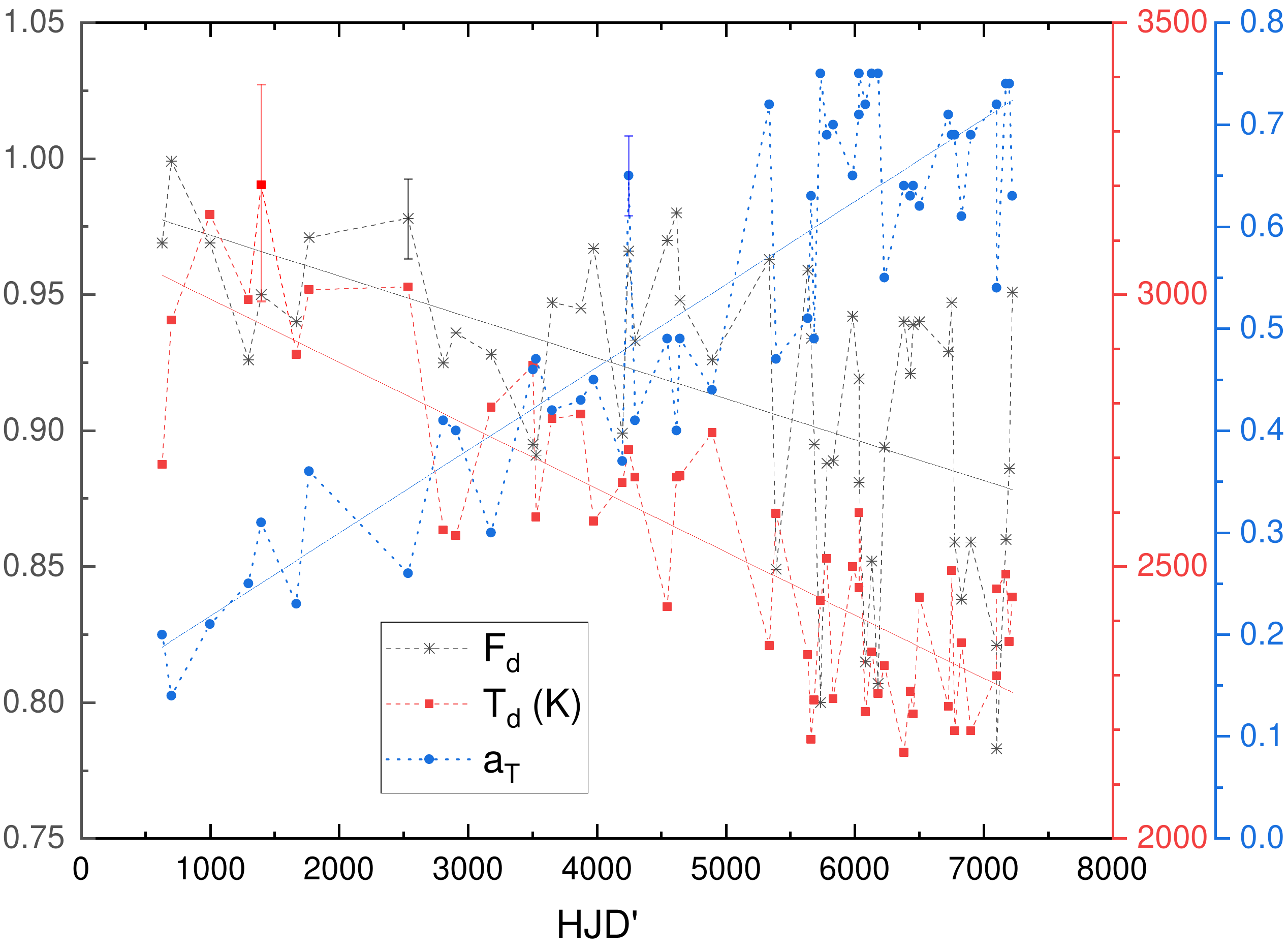}}
\caption{Results of the PCA analysis along with the time evolution of some disk parameters. Left panel: PCA for the parameters given in Table \ref{tab:data}.  The two principal components, PC1 and PC2, result from lowering the dimension of the dataset.
Some parameters are shown in color for a better visualization. 
Right panel: Temporal evolution of the disk filling factor, its temperature, and temperature exponent.  Best fits are also shown, with their parameters given in Table \ref{tab:paramfits}.  Typical uncertainties are 0.03, 200 K, and 0.03 for $F_d$, $T_d$ and $a_T$, respectively. They are shown for representative points to avoid data crowding.}
\label{fig:PCA}
\end{figure*}

The result of the PCA analysis is shown in Fig. \ref{fig:PCA} left. The PC1 and PC2 components capture most of the data variability, namely, 30.14\% and 18.46\%, respectively.  
 The linear decompositions of PC1 and PC2 in terms of the studied parameters are given in Table \ref{tab:pcatable}.
The clustering of parameters in the graph indicates correlation (or an anticorrelation if they are in opposite quadrants) and allows us to draw the following conclusions:

 There are (anti)correlations between $HJD'$ and the parameters of temperature, disc radial extension, and temperature index $(T_d$, $R_d$, and $a_T$). This can be interpreted in terms of the link between these quantities through Eq.\,2, along with the tendency of decreasing disk radius with HJD' that has already been noted. 

There are correlations between the magnitude at orbital phase ($\Phi_o$) equal to 0.25, $I_{0.25}$, and the thicknesses of the disk, $d_e$ and $d_c$. This is interesting since it links the long cycle photometric variability with a disk variable parameter and suggests a direct cause for the long cycle. The magnitude at orbital phase 0.25, $I_{0.25}$, also is anticorrelated with the hot spot position,  $\theta_{hs}$, and  to a lesser degree with the bright spot size, $\theta_{bs}$.
The positions of the hot and bright spots, $\lambda_{hs}$ and $\lambda_{bs}$, are anticorrelated.
The relative temperature of the hot spot $A_{hs}$ is anticorrelated with its angular size $\theta_{hs}$. The mass transfer rate, $\dot{M,}$  is correlated with the position of the bright spot $\lambda_{bs}$ and anticorrelated with the position of the hot spot $\lambda_{hs}$.

\begin{figure}
\scalebox{1}[1]{\includegraphics[angle=0,width=8.5cm]{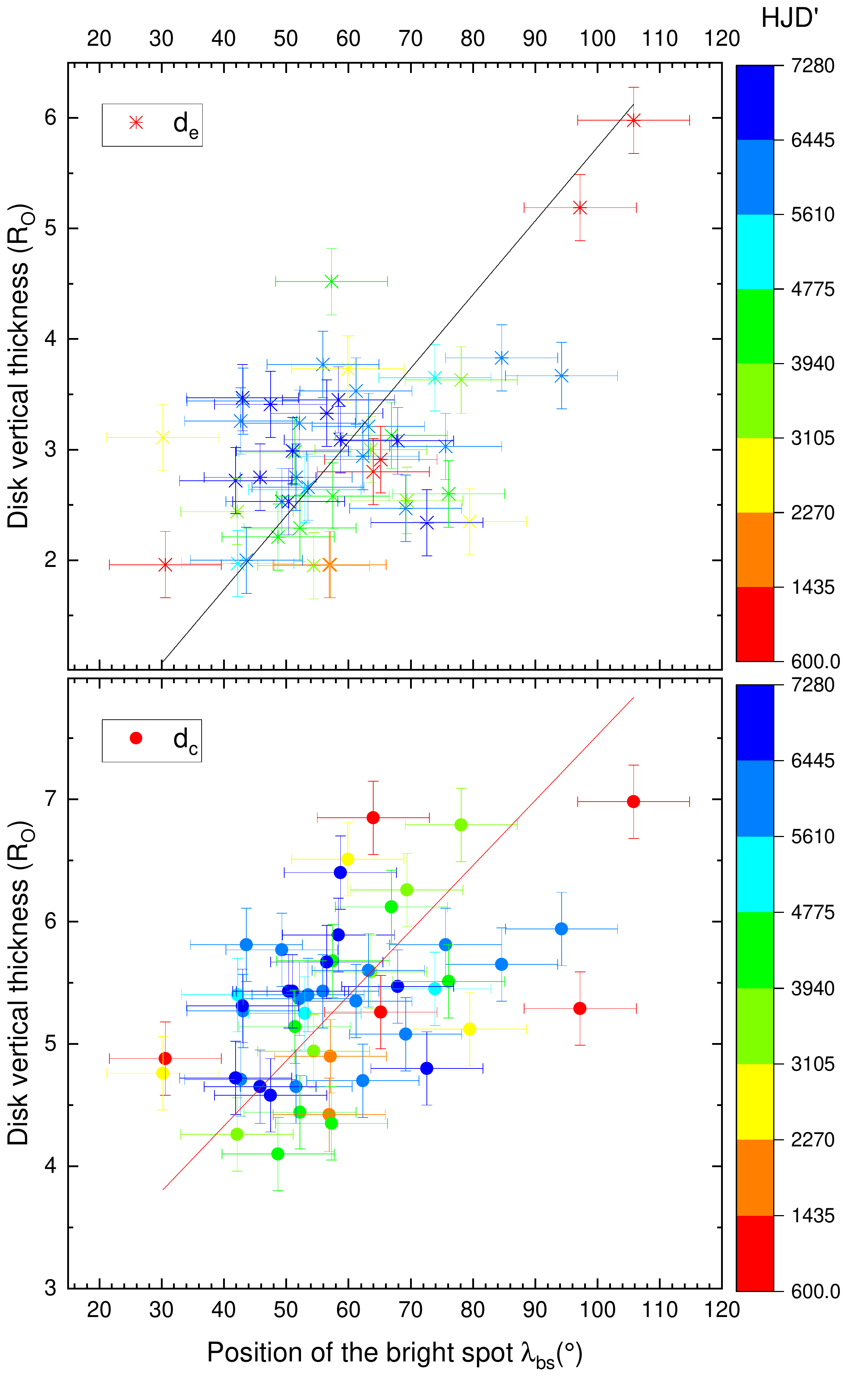}}
\caption{Disk vertical thickness  versus position of the bright spot. Best  linear fits are also shown, with their parameters
given in Table \ref{tab:paramfits}.}
\label{fig:brightspot}
\end{figure}

\begin{figure}
\scalebox{1}[1]{\includegraphics[angle=0,width=8.5cm]{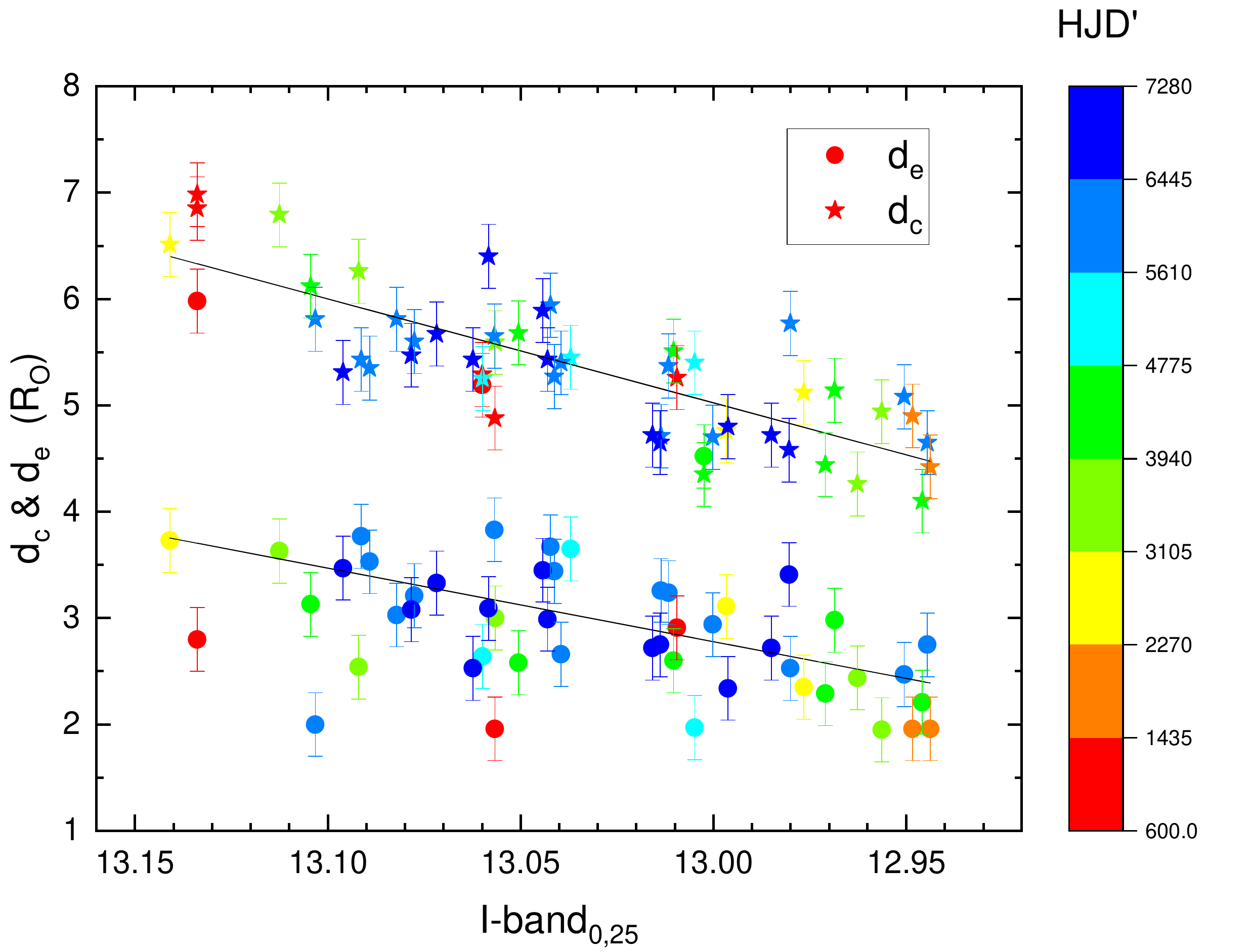}}
\caption{Dependence of disk vertical thickness with the system magnitude at  $\Phi_o$=  0.25. Best  linear fits are also shown, with their parameters given in Table \ref{tab:paramfits}.} 
\label{fig:dedcI025}
\end{figure}

\begin{figure*}
\scalebox{1}[1]{\includegraphics[angle=0,width=18cm]{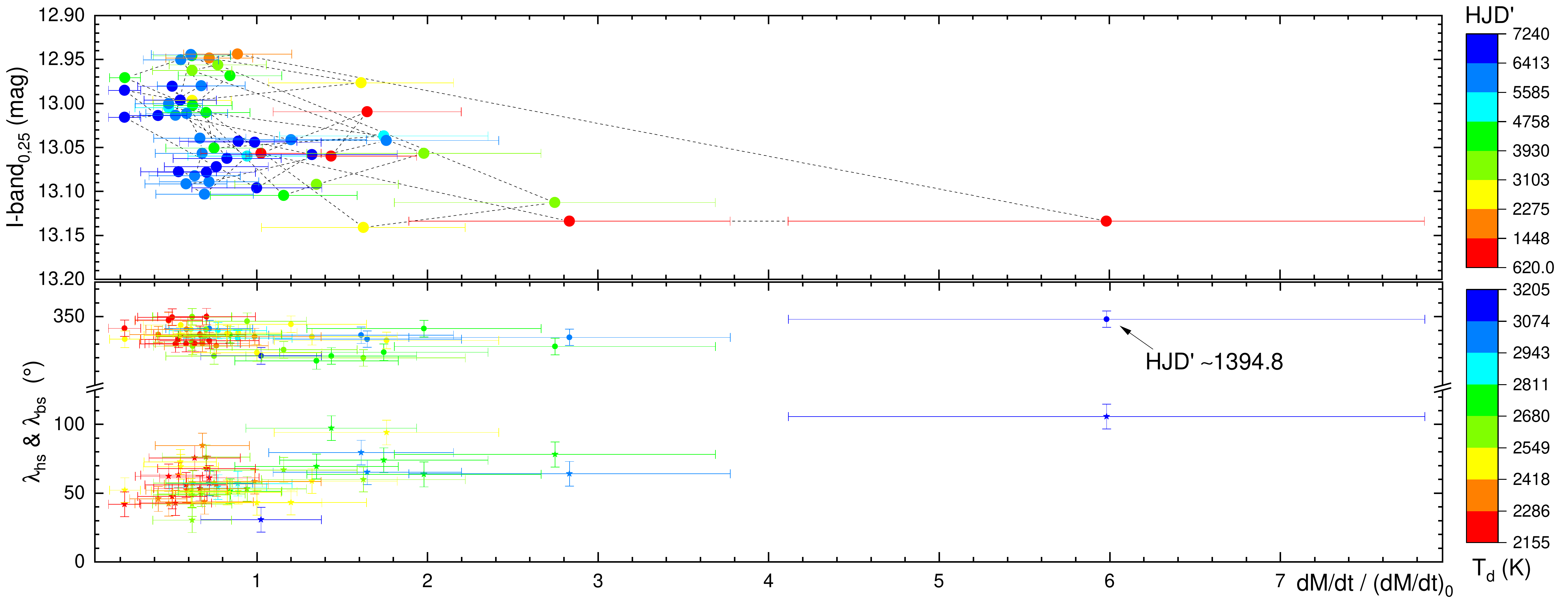}}
\caption{System magnitude at $\Phi_o$= 0.25 and spot positions versus normalized mass transfer rate.  In the upper panel,  dashed lines join consecutive data points.}
\label{fig:dmdt}
\end{figure*}

\begin{figure}
\scalebox{1}[1]{\includegraphics[angle=0,width=8.5cm]{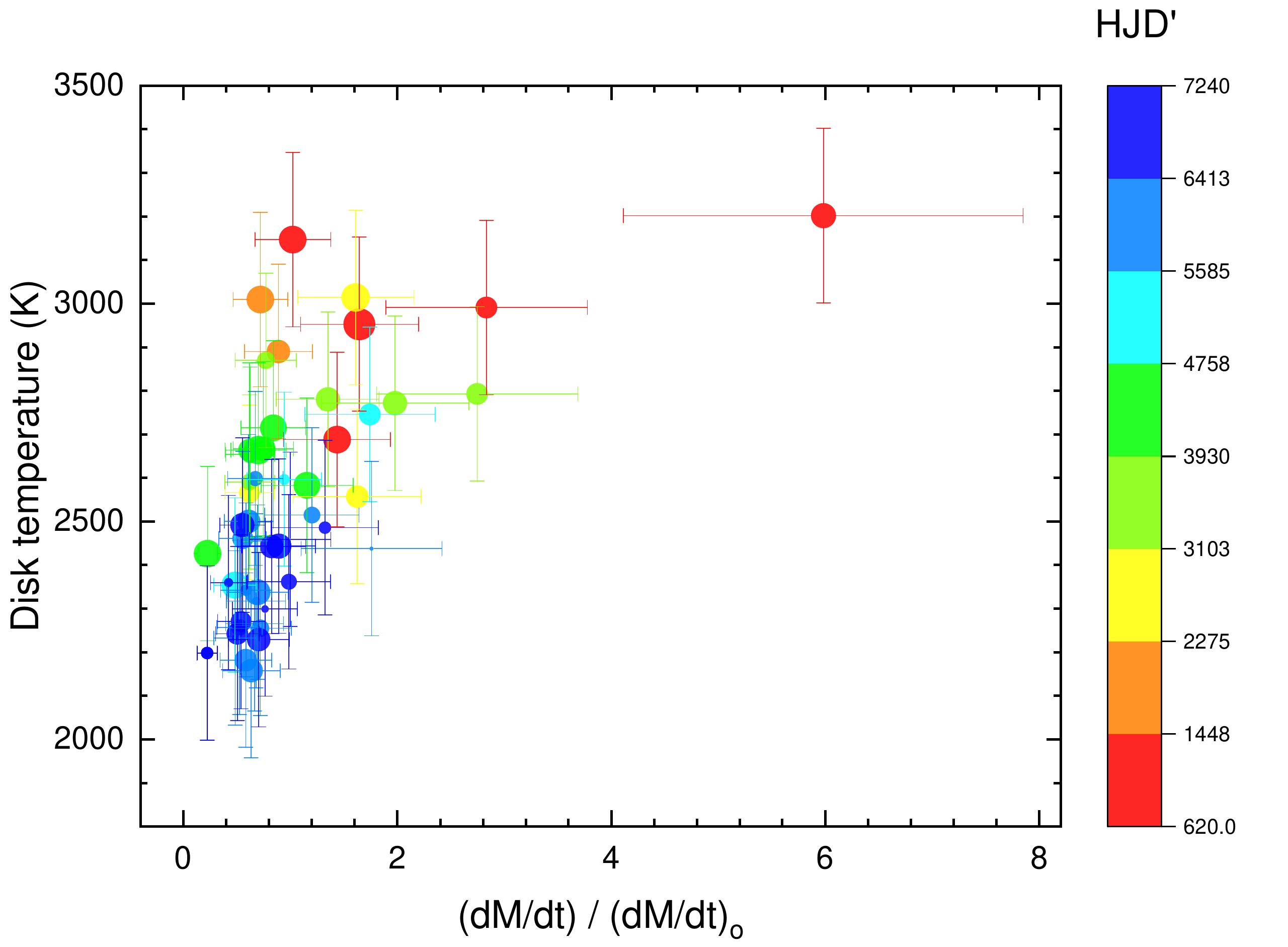}}
\caption{Mass transfer rate versus disk temperature. The colors represent HJD' ranges.  Symbol size scales with disk radius.} 
\label{fig:dotMT}
\end{figure}

\begin{figure}
\scalebox{1}[1]{\includegraphics[angle=0,width=8.5cm]{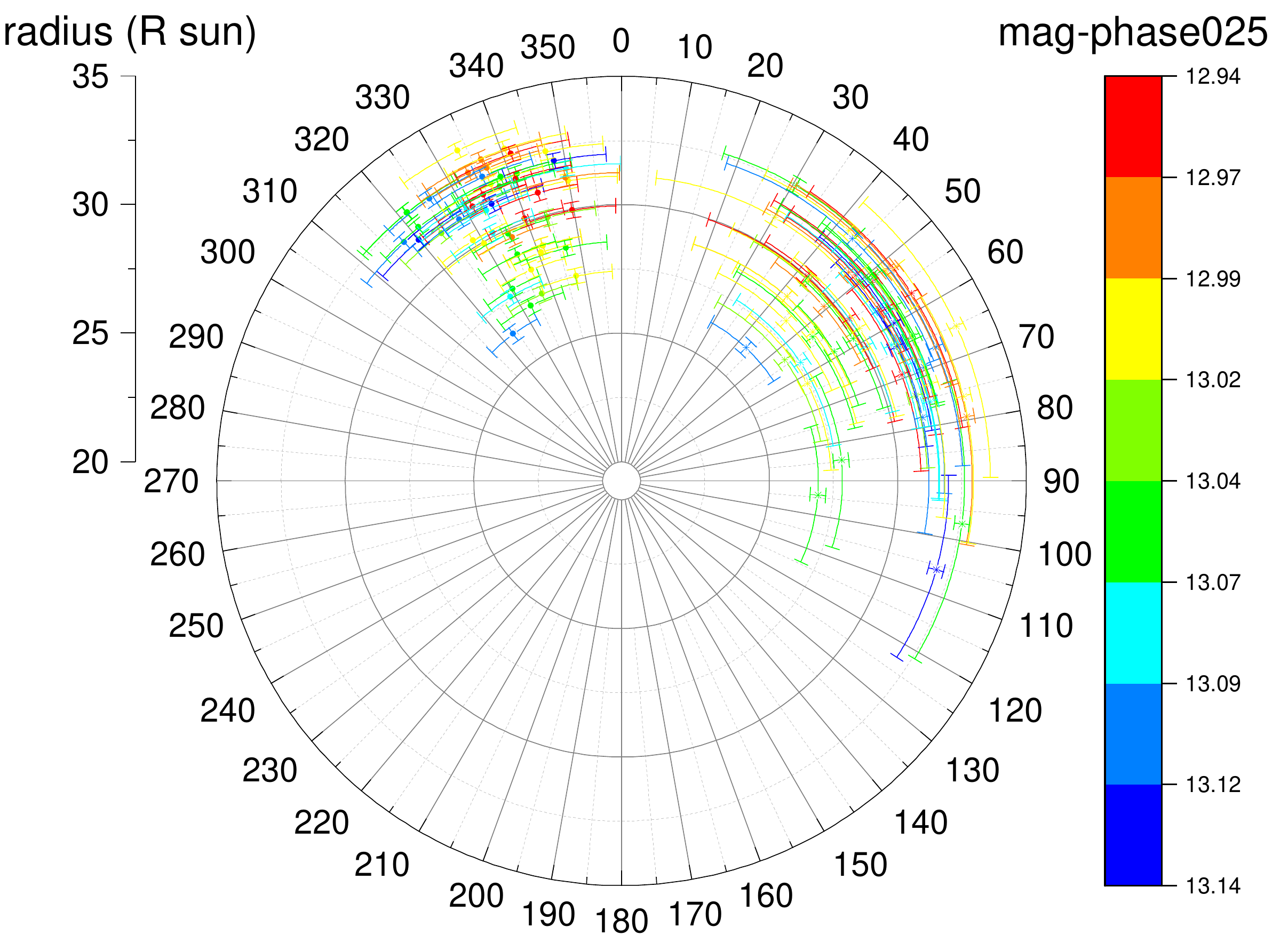}}
\caption{Radial and angular distribution of hot and bright spots based on the $R_d$, $\theta,$ and $\lambda$ parameters.  Errors in angular position and size for the hot and bright spots are 2\fdg5, 6$^\circ$ and 8$^\circ$, 9$^\circ$, respectively. The inner and outer circles represent radii of 20 and 35 R$_{\odot}$. }
\label{fig:angular}
\end{figure}

The above results can be evaluated in the plots shown in  Figs. \ref{fig:PCA} (right) and Figs. \ref{fig:brightspot} to \ref{fig:angular}. We have made linear fits in some cases, of the form $y = a + b x$, whose parameters are given in Table \ref{tab:paramfits}. Color bars in the above plots indicate HJD' or temperature ranges.
From these plots, the following conclusions can be outlined: 

The central and edge vertical thickness of the disk increases with the angular position of the bright spot (Fig. \ref{fig:brightspot}).
When the system is fainter, the disk is thicker at its center and its outer edge (Fig. \ref{fig:dedcI025})

Larger mass transfer rate corresponds to fainter system brightness at phase 0.25. When time goes on and the disk size decreases, this relationship becomes steeper,  as becomes evident noticing the colored data in  the upper panel of Fig. \ref{fig:dmdt}.  In addition, the mass transfer rate attains a maximum around HJD' =  1394.8, namely, at epochs when the system shows larger amplitude of the long-cycle and a larger disk.

When mass transfer rate increases, the hot and bright spots becomes closer to the donor (Fig. \ref{fig:dmdt}, lower panel).
The mass transfer rate increases with disk temperature (Fig. \ref{fig:dotMT}).

The hot spot shows a smaller variability in position compared with the bright spot. Both are located in the disk edge, at opposite sides of the donor star. No dependence is observed between disk radius and spot position (Fig. \ref{fig:angular}). The disk tends to be larger at brighter magnitudes at orbital phase 0.25 (Fig. \ref{fig:angular}).

\begin{table}
\centering
\caption{ Coefficients of the linear decomposition of principal components PC1 and PC2.}
\label{tab:pcatable}
\begin{tabular}{lrrlrr} 
\hline
Par. & PC1 &PC2 &Par.  &PC1 &PC2  \\
\hline
$HJD'$& -0.35992&       0.32759 &                       $\theta_{rad}$& 0.00011&        -0.23296\\         
$T_d$ & 0.35045&        -0.26502&$A_{bs}$ & -0.33308           &        0.01494\\
$d_e$ & 0.22398&        0.32982& $\theta_{bs}$ & -0.05614    &  -0.22559\\
$d_c$ & 0.26740  &     0.37307&$\lambda_{bs}$ & 0.29367&      0.17763\\
$a_d$& -0.35849             &   0.28694&$R_d$ & 0.22041   &  -0.30856\\
$A_{hs}$  & -0.06673        &   0.17510&$I_{0.25}$ & 0.23524&   0.40123\\
$\theta_{hs}$ & 0.09600      &  -0.16992&$\dot{M}$& 0.38702        &    0.15727\\
$\lambda_{hs}$ & -0.17248&      -0.12143& & & \\
\hline
\end{tabular}
\end{table}

\begin{table}
\centering
\caption{Parameters of the fits of type $y = a + b x$. Pearson's $r$ parameter is given, along with the errors of the coefficients.}
\label{tab:paramfits}
\begin{tabular}{lrrrrr} 
\hline
Dataset & $a$ &$\epsilon_a$ &$b$ &$\epsilon_b$ &$r$ \\
\hline
$HJD'$  $a_T$ &0.1371&0.0298 &8.13E-5&5.7E-6&0.90 \\
$HJD'$  $T_d$ &3108.3&55.8&-0.1162&0.0107&-0.84\\
$HJD'$  $F_d$ &0.9869&0.0166&-1.50E-5&3.2E-6&-0.56\\
$I_{0.25}$  $d_e$ &-87.02       &23.13 &6.91&   1.78 &0.48 \\
$I_{0.25}$  $d_c$ &-121.94&     13.62 &9.77     &1.05& 0.80 \\
$\lambda_{bs}$  $d_c$ &2.20&    0.55& 0.0533&    0.0091& 0.50\\
$\lambda_{bs}$  $d_e$ &-0.94&0.67 &0.0667       &0.0011& 0.55\\
$T_d$  $R_d$ & 22.50&   2.13 &0.0032&   0.0008 & 0.48\\
\hline
\end{tabular}
\end{table}

\begin{figure}
\scalebox{1}[1]{\includegraphics[angle=0,width=8.5cm]{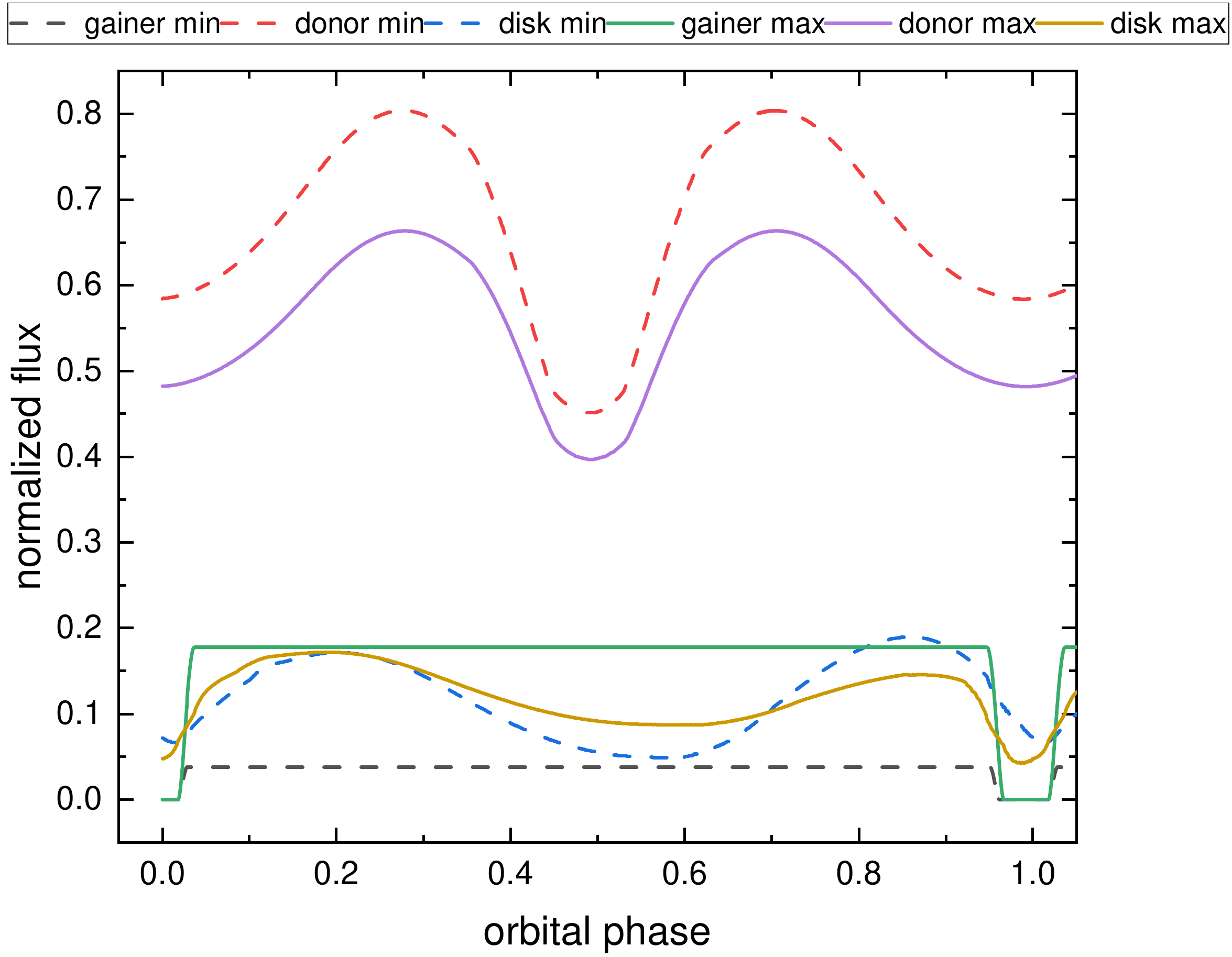}}
\caption{Fluxes at the $I$-band of gainer, donor and disk per orbital phase, normalized to the total flux at every orbital phase. 
Data at the two epochs for the minimum (dashed lines) and maximum (solid lines) of the long cycle reported by \citet{2020A&A...641A..91M} are given. These data are the datasets labeled 11 and 44, respectively, in Table \ref{tab:data}.}
\label{fig:fluxes}
\end{figure}

\begin{figure*}
\scalebox{1}[1]{\includegraphics[angle=0,width=18cm]{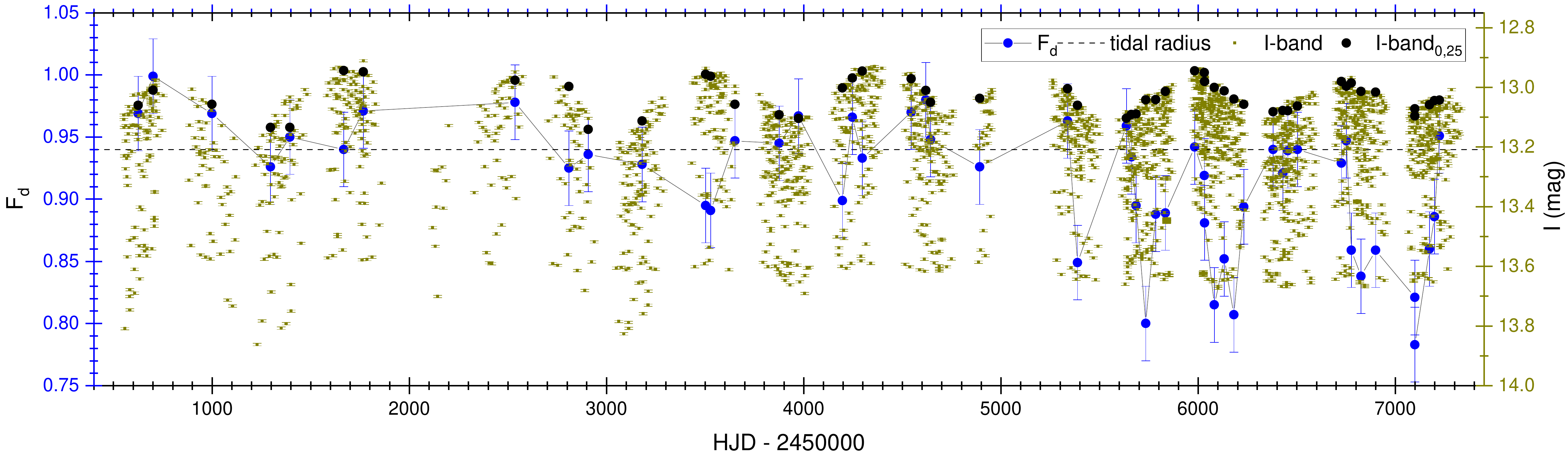}}
\caption{Fractional disk radius, $I$-band magnitude, and $I$-band at orbital phase 0.25. This last one traces the DPV long-cycle. The tidal radius is indicated by an 
horizontal dashed line.}
\label{fig:tidal}
\end{figure*}

\section{Discussion}

Before  continuing  with the analysis and interpretation of our results, we must keep in mind the limitations of our model (some of them already mentioned in Section 3.1). The main approximation is probably based on only 
the circumstellar material in the disk being considered, neglecting any light contribution with an origin located outside this plane above or below the disk. An additional approximation is the assumption of a disk shape for the main circumstellar structure. With our model, we do not consider other possible light sources like jets, winds, and outflows.  
However, the two disk spots can, in principle, take into account variations in the  azimuthally dependent disk emissivity. In spite of the aforementioned limitation, and based on previous studies of Algols with disks, we have probably considered the main light sources for the $I$-band  continuum light, as reflected in the very good match of orbital and long-term light curves over the 18.5 years of observations (Fig. \ref{fig:fits}).

During the 18.5 year baseline, the system shows a systematic decrease in disk size and temperature. At the beginning, the temperature was about 3000 $K$ and the disk almost filled the Roche lobe of the late B-type star in the radial coordinate. At the end of the time series, the temperature dropped to 2300 $K$ and the disk filled about 85\% of the Roche lobe radial extension. This very long-term tendency was accompanied with a large increase in the temperature power-law index. Remarkably, there are additional changes in disk radius and temperature around these general long-term mean tendencies. These faster changes are more on timescales of hundreds of days and are associated with the DPV long cycle of this binary.

The brightness of the system at orbital phase 0.25  turns to be a useful indicator of the DPV long cycle  and correlates with disk vertical thickness, so the system is brighter when the disk is thinner. 
This explains the DPV long cycle in terms of differential occultation of the hot star. This occultation turns out to be clear when comparing the relative flux contribution of donor, disk, and gainer to the total flux during bright and faint states  (Fig. \ref{fig:fluxes}). 

The brightness also is anticorrelated with the mass transfer rate. 
Larger mass transfer rates produce a thicker and hotter disk along with a fainter system. We conclude that 
the variable mass transfer rate modulates the overall photometric variability. When the disk is larger, it is also thicker and hotter, and the system is fainter. On the other hand, when the system brighter, it  shows disk spots closer to the donor.  This last finding could be understood in terms of larger turbulent motions during the thicker and hotter disk stage associated with a larger mass transfer rate, which might help to distribute the heat from the spots to the surrounding disk regions. This behaviour could be tested in the future with hydrodynamical simulations. In addition, the lack of correlation between disk temperature and outer disk thickness probably indicates that the disk is not in hydrostatic equilibrium and turbulent motions are present. In this context, it may be worth mentioning that in a detailed model of $\beta$ Lyrae, the disk height has to be multiplied  by a factor about four times its expected equilibrium value, possibly reflecting non-negligible hydrodynamic flows within the disk \citep{2021A&A...645A..51B}.

In hydrostatic equilibrium, the vertical height, $H,$ of an accretion disk at radius, $R_d$, is:
\begin{equation}
\frac{H}{R_d} \approx \frac{c_s}{v_k} = c_s \sqrt{\frac{R_d}{GM_1}},
\end{equation}

\noindent
where $v_k$ is the Keplerian velocity and $c_s$ the sound speed, which, for an isothermal perfect gas, can be approximated as:\\
 \begin{equation}
c_s \approx 10 \sqrt{\frac{T}{10^4\,K}} \, \rm{km\,s^{-1}},
\end{equation}

\noindent
\citep[Eqs. 3.35 and 3.32 in][]{2010eea..book.....K}.
Using the mean parameters at the outer and inner disk we get ($v_k,c_s$) = (172.8, 5.0) and (453.2, 11.8) in km\,s$^{-1}$, respectively, yielding $H/R$ = 0.029 and 0.026.  Considering the averages $d_e/R_d$ = 0.098 and $d_c/R_1$ = 1.19 and that the vertical thickness is twice $H$, we conclude that the disk vertical height is larger than expected for hydrodynamical equilibrium, at the inner and outer boundaries, reinforcing the idea that turbulent motions dominate the disk vertical structure. 

In our model, the maximum possible disk radius is 33.64  $R_{\odot}$, corresponding to $F_d$ = 1.0. Actually, this size is attained around HJD'= 682.53 at a time when the amplitude of the long cycle is the largest. In principle, the disk can be stable until the last non-intersecting orbit defined by the tidal radius \citep[][Eq. 2.61]{1977ApJ...216..822P, 1995CAS....28.....W}:
\begin{eqnarray}
\frac{r_{t}}{a_{orb}}= \frac{0.6}{1 + q},
\end{eqnarray}

\noindent we get $r_t/a_{orb}$ = 0.492, or $r_t$ = 31.77 $R_{\odot}$.  We observe that  during all the observing epochs, the disk shows cycles of radial extension around the tidal radius (Fig. \ref{fig:tidal}). 
We also observe that the disk radius changes are of smaller amplitude when the disk is large and above the tidal radius, whereas they are of larger amplitude when the disk is smaller and usually below the tidal radius.

The disk radius oscillates around the tidal radius rather erratically regarding the long DPV cycle traced by the system brightness, although  it takes place on a similar timescale.   A possible explanation for the above behavior is that the disk reacts rather rapidly to vertical transport of mass, adapting its structure (in particular, the vertical thickness) to changing mass transfer rates. This should happen on a dynamical timescale. However, the radial extension is controlled by the much longer viscous timescale, hence, the disk radius changes more slowly and with a delay.  A likely supporting fact for this picture is the decrease in disk radius observed on a decade-long timescale following a sustained decrease in the mass transfer rate.   It is worth mentioning that when time evolves and the disk external radius drops below the tidal radius, $a_T$ goes to 0.75, hence, the disk evolves to the steady state condition (Fig.\,\ref{fig:PCA}, right).

The linear theory of tidal interaction of an accretion disk in a close binary systems predicts the generation of spiral waves at regions characterized by a Lindblad resonance.
A $j:k$ Lindblad resonance occurs when $j$ times the disk angular speed $\Omega$ is commensurable with $k$ times the orbital angular speed $\omega$. 
This happens when \citep{1991MNRAS.249...25W}:
\begin{eqnarray}
\frac{r_{jk}}{a_{orb}} = \left( \frac{j-k}{j} \right) ^{2/3} (1+q)^{-1/3}.
\end{eqnarray}

It has been shown that 3:1 and 2:1 resonances play a role in the accretion disks of cataclysmic variables, explaining phenomena such as superhumps observed in the light curves during 
superoutbursts of the SU\,UMa type systems in terms of disk precession \citep{1991MNRAS.249...25W}.  Disk precession is rarely mentioned in the literature on Algols, although has been suggested as possible cause for the 50 day period found in radio data of $\beta$ Per \citep{2005ApJ...621..417R}.
For our system, we get $r_{2:1}/a_{orb}$ =  0.59 and $r_{3:1}/a_{orb}$= 0.71 
or  38.11 and 45.87 
$R_{\odot}$.  Both resonance radii exceed the disk size and, therefore, they do not play a role in the phenomena observed in this system. It is,  in principle, possible that the influence of the 2:1 resonance expands beyond the 2:1 radius, but it is hard to imagine that it does so even when the disk is smaller than the tidal radius when the phenomenon is still visible. Hence, while in SU\,UMa stars the light curve oscillations are due to enhanced energy dissipation in precessing disks extending to the resonance radius, the long-term light oscillations observed in \var\ cannot possibly have the same origin and are probably caused, as stated before, by the occultation of the hot star by a thick disk. 

The discovery that a disk of variable thickness is very likely  responsible for the DPV long cycle in \var\ poses a challenge to previous interpretations for such a cycle, 
such as a variable bipolar bipolar wind \citep{2012MNRAS.427..607M}
or an ejected circumbinary disk \citep{2008MNRAS.389.1605M}. The above constrains were based on the study of line emissions and absorptions, probably reflecting the variability of optically thin disk regions.  These kinds of regions might co-exist with the optically thick disk in $\beta$ Lyrae type binaries \citep{2021A&A...645A..51B}. On the contrary, our model is sensible to the optically  thick disk, contributing to the continuum fluxes emitted in the $I$-band. We conjecture that the previous interpretations are complementary. A larger mass transfer rate not only affects the optically thick disk but also the wind emanating from the hotspot region or even outflows escaping from the system through the $L_3$ point when the disk is large enough. It is reasonable to assume now that these line emission phenomena associated with the long-cycle are not the cause for the light curve long cycle, but additional phenomena related to the cause of disk thickness change,  for example,   
a variable mass transfer rate. 
 
Regarding the origin of the mass transfer changes, we have not found any new information from the analysis of the light curve. As stated by \citet{2020A&A...641A..91M}, the long-term cycle length fits the magnetic dynamo model proposed by \citet{2017A&A...602A.109S}. However, further spectroscopic and polarimetric studies are needed to explore the possible magnetic nature of the donor star.

 Our finding of disk structural changes during the long cycle are in line with those reported for the DPV OGLE-LMC-DPV-097 \citep{Garces2018}. For this system, the rapid and large amplitude changes associated with the long cycle have not permitted an investigation of the orbital light curve during the whole time baseline and the study focused on the mean high and low stages only. The above suggests that disk structural changes might be a characteristic of many (if not all) DPVs.

\section{Conclusion}

In this work, we investigate the interesting light curve of the $\beta$ Lyrae type and double periodic variable \var, spanning 18.5 years of $I$-band data. We constructed a model for the system that is able to reproduce the overall light curve at three timescales: the orbital one of 24\fd8, the DPV cycle of hundreds of days, and the overall long-term tendency that spans decades. Our main results, which are new and serve as a complement to those reported in the seminal work by  \citet{2020A&A...641A..91M}, can be summarized as follows:\\

\begin{itemize}

\item The long super-orbital photometric changes can be understood as changes in the mass transfer rate of the system, $\dot{M}$. 
\item These changes  occur on a timescale of hundreds of days and correspond to the reported  DPV cycle, but also happen on a decade-length timescale. 
\item   The disk radius cyclically changes around the tidal radius on a timescale of hundreds of days, decoupled of changes in $\dot{M}$  or system brightness.
\item  On the contrary, on the decade length time scale the disk radius  follows $\dot{M}$, which might be explained in terms of a slow viscous response to mass input.
\item The decade-length timescale variability is characterized by a decrease in disk radius and disk temperature, along with a decrease in  $\dot{M}$.  The disk temperature index $a_T$ goes to 0.75, namely to the steady state condition, when the time evolves and the disk external radius drops below the tidal radius.
\item When $\dot{M}$ is large, the disk thickness increases, as  does its temperature. At the same time the system brightness decreases due to the  occultation of the gainer by a thicker disk.
\item  In our model, the DPV cycle is produced by time-dependent occultation of the gainer by a disk of variable thickness.
\item  Lindblad resonance regions are far beyond the disk radius, excluding viscous dissipation or disk precession as a major source of photometric variability.
\end{itemize}

\begin{acknowledgements}
 We thanks the referee Dr. Miroslav Broz for his comments that contributed to improve the first version of this manuscript. 
REM acknowledges support by BASAL Centro de Astrof{\'{i}}sica y Tecnolog{\'{i}}as Afines (CATA) PFB--06/2007 and FONDECYT 1190621. 
G.D. acknowledges the financial support the Ministry of Education, Science and Technological
Development of the Republic of Serbia through contract No. 451-03-68/2020-14/200002. Thanks to Daniela Francisca Mennickent Barros for calling our attention to the PCA analysis tool. 
Thanks to Juan Garc\'es Letelier for his comments on earlier versions of this manuscript. 
\end{acknowledgements}

 \begin{appendix}
\clearpage
\section{Additional tables and figures}

\begin{figure*}
   \begin{center} 
\scalebox{1}[1]{\includegraphics[angle=0,width=16cm]{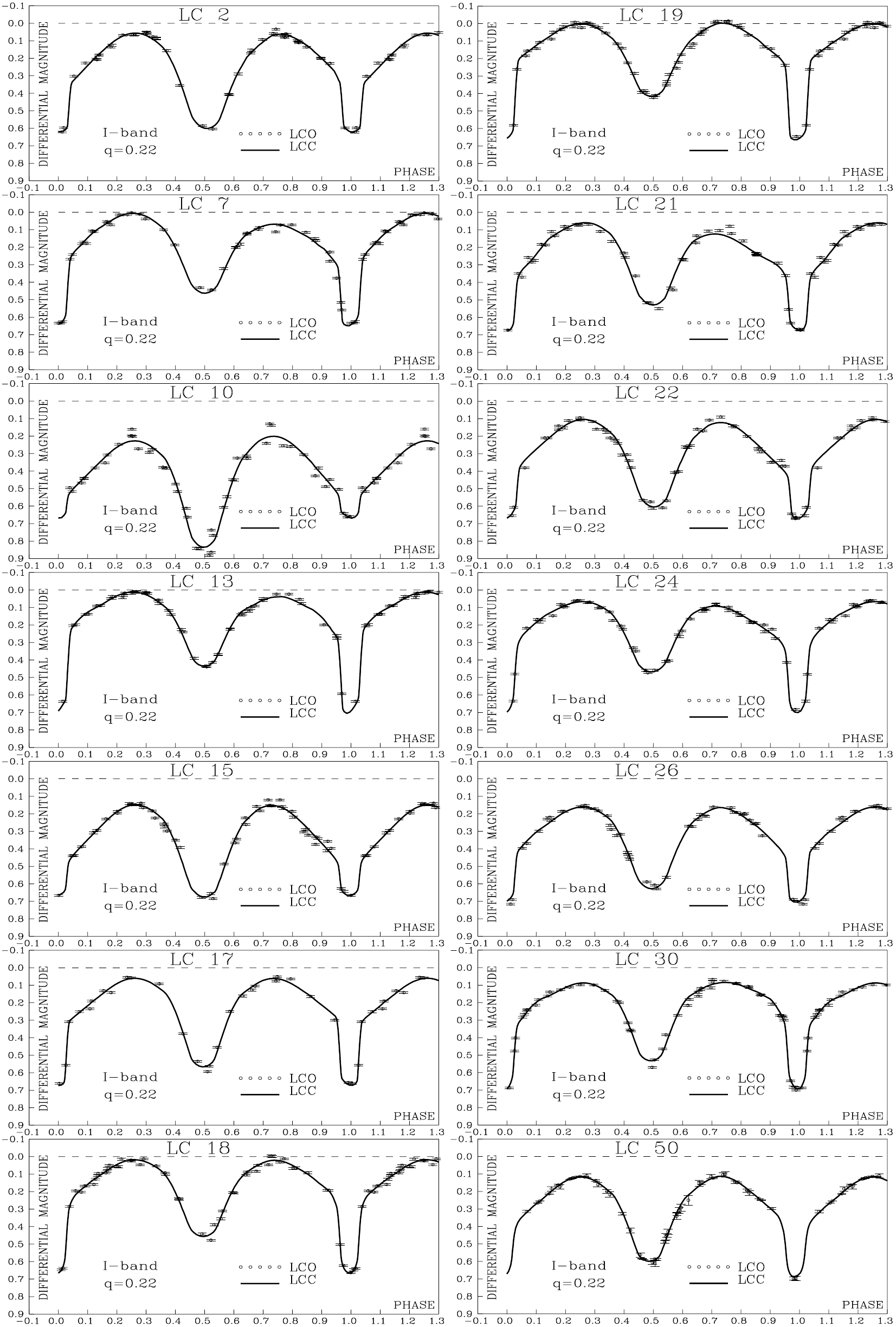}}
\caption{Orbital $I$-band differential light curves and the respective fits for some of the data segments listed in Table \ref{tab:data}. The dashed line indicates the maximum brightness of the system in the orbital phase 0.25 during all the observations. All the brightness curves are normalized in relation to this maximum brightness.}
\label{fig:fits}
\end{center}
\end{figure*}

\begin{table*}
\centering
\caption{Light curve fit parameters for data segments characterized by label LC and average $HJD-2450000$ $(HJD`)$. }
\label{tab:data}
\begin{tabularx}{1.0\textwidth}{
>{\raggedleft\arraybackslash}X  
>{\raggedleft\arraybackslash}X 
>{\raggedright\arraybackslash}X
>{\raggedright\arraybackslash}X  
>{\centering\arraybackslash}X 
>{\raggedleft\arraybackslash}X
>{\raggedright\arraybackslash}X  
>{\centering\arraybackslash}X 
>{\raggedright\arraybackslash}X
>{\raggedright\arraybackslash}X  
>{\centering\arraybackslash}X 
>{\raggedleft\arraybackslash}X
>{\raggedright\arraybackslash}X  
>{\centering\arraybackslash}X 
>{\raggedleft\arraybackslash}X
>{\raggedright\arraybackslash}X  
>{\centering\arraybackslash}X }
\hline
LC&     $HJD'$  &$F_{d}$        &$T_{d}$&       $a_{T}$&        $A_{hs}$&       $\theta_{hs}$&  $\lambda_{hs}$& $\theta_{rad}$& $A_{bs}$&       $\theta_{bs}$&  $\lambda_{bs}$& $R_{d}$ &$d_{e}$&       $d_{c}$&        $I_{0.25}$&     $\dot{M}/\dot{M}_0$ \\
 &  & $\pm$0.03&$\pm$200&$\pm$0.03&$\pm$0.05 & $\pm$2.5 &$\pm$6.0 &$\pm$7.0 &$\pm$0.04&$\pm$8.0&$\pm$9.0 &$\pm$0.3&$\pm$0.3&$\pm$0.3&$\pm$0.01&  $\pm$50\% \\
 & & & (K) & & &($^{o}$) &($^{o}$) &($^{o}$) & & ($^{o}$) &($^{o}$) &(R$_{\odot}$) &(R$_{\odot}$) &(R$_{\odot}$) & (mag) & \\
\hline
1&       625.8  &0.97&  2688    &0.20&  1.67&   20.4&   321.3&  0.6     &1.34&  48.1&   97.2&   32.60&  5.19&   5.29&   13.06&  1.44\\
2&       700.5  &1.00&  2953    &0.14&  1.84&   19.6&   333.6&  $-$24.6 &1.50&  48.5&   65.2&   33.60&  2.91&   5.26&   13.01&  1.65\\
3&       998.6  &0.97&  3147    &0.21&  1.78&   18.8&   321.4&  $-$20.9 &1.54&  26.6&   30.6&   32.61&  1.96&   4.88&   13.06&  1.02\\
4&      1295.9  &0.93&  2991    &0.25&  1.83&   20.2&   334.9&  $-$20.6 &1.56&  39.1&   64.0&   31.15&  2.80&   6.85&   13.13&  2.83\\
5&      1394.8  &0.95&  3202    &0.31&  1.73&   18.3&   348.1&  $-$17.2 &1.34&  33.7&105.8             &31.97&     5.98& 6.98&   13.13&  5.98\\
6&      1667.8  &0.94&  2890    &0.23&  1.96&   19.0&   334.2&  $-$6.2  &1.69&  49.5&   56.9&   31.61&  1.96&   4.42&   12.94&  0.89\\
7&      1766.5  &0.97&  3009    &0.36&  1.68&   19.8&   341.3&  $-$13.6 &1.62&  48.0&   57.1&   32.68&  1.96&   4.90&   12.95&  0.72\\
8&      2535.6  &0.98&  3014    &0.26&  1.90&   20.5&   336.4&  $-$28.9 &1.55&  42.0&   79.5&   32.90&  2.35&   5.12&   12.98&  1.61\\
9&      2808.8  &0.93&  2567    &0.41&  1.74&   18.5&   349.9&  1.5     &1.85&  47.3&   30.2&   31.11&  3.11&   4.76&   13.00&  0.62\\
10&     2907.6& 0.94&   2557    &0.40&  1.81&   18.6    &319.9& $-$23.9 &1.50&  41.7&   59.9    &31.50  &3.73   &6.51   &13.14  &1.62\\
11&     3180.6& 0.93&   2793    &0.30&  1.84&   19.7    &328.2& $-$12.9 &1.84&  43.4&   78.1    &31.21  &3.63   &6.79   &13.11  &2.75\\
12&     3502.9& 0.90&   2870    &0.46&  1.87&   16.4    &339.7& 3.6     &1.66&  46.7&   54.4    &30.13  &1.95   &4.94   &12.96  &0.77\\
13&     3527.8& 0.89&   2591    &0.47&  1.94&   18.2    &349.7& $-$27.1 &1.75&  47.6&   42.1    &29.96  &2.44   &4.26   &12.96  &0.62\\
14&     3651.5& 0.95&   2772    &0.42&  1.98&   19.6    &341.2& $-$27.0 &1.61&  26.2&   63.6    &31.87  &3.00   &5.59   &13.06  &1.98\\
15&     3874.8& 0.95&   2780    &0.43&  1.76&   19.9    &317.7& $-$25.6 &1.53&  45.6&   69.4    &31.81  &2.54   &6.26   &13.09  &1.35\\
16&     3973.6& 0.97&   2583    &0.45&  1.81&   17.4    &325.8& $-$16.8 &1.54&  41.1&   66.9    &32.53  &3.13   &6.12   &13.10  &1.16\\
17&     4196.8& 0.90&   2654    &0.37&  1.68&   15.4    &328.3& $-$15.0 &1.40&  39.1&   57.3    &30.24  &4.52   &4.35   &13.00  &0.62\\
18&     4246.8& 0.97&   2715    &0.65&  1.73&   18.4    &336.7& $-$20.7 &1.46&  43.7&   51.4    &32.50  &2.98   &5.14   &12.97  &0.84\\
19&     4296.6& 0.93&   2664    &0.41&  1.90&   21.3    &331.5& $-$26.2 &1.78&  33.2&   48.7    &31.38  &2.21   &4.10   &12.95  &0.62\\
20&     4544.9& 0.97&   2426    &0.49&  1.61&   18.2    &333.5& 21.3    &1.82&  43.3&   52.2    &32.63  &2.29   &4.44   &12.97  &0.23\\
21&     4618.7& 0.98&   2664    &0.40&  1.64&   20.5    &340.6& 20.5    &1.51&  48.2&   76.1    &32.95  &2.60   &5.51   &13.01  &0.70\\
22&     4643.6& 0.95&   2667    &0.49&  1.75&   15.8    &321.3& $-$29.8 &1.67&  37.3&   57.5    &31.89  &2.58   &5.68   &13.05  &0.75\\
23&     4891.9& 0.93&   2746    &0.44&  1.91&   17.9    &324.0& $-$12.0 &1.62&  27.4&   73.9    &31.15  &3.65   &5.45   &13.04  &1.74\\
24&     5337.8& 0.96&   2354    &0.72&  1.84&   19.6    &347.0& 5.7     &1.79&  34.2&   42.2    &32.40  &1.97   &5.40   &13.00  &0.48\\
25&     5387.8& 0.85&   2597    &0.47&  1.86&   19.8    &346.6& 7.0     &1.53&  48.4&   53.0    &28.56  &2.64   &5.25   &13.06  &0.94\\
26&     5362.5& 0.96&   2338    &0.51&  2.00&   17.9    &335.4& $-$29.9 &1.78&  50.7&   43.6    &32.26  &2.00   &5.81   &13.10  &0.69\\
27&     5660.8& 0.93&   2182    &0.63&  1.76&   20.3    &330.3& $-$16.4 &1.49&  48.0&   55.9    &31.41  &3.77   &5.43   &13.09  &0.58\\
28&     5685.3& 0.90&   2255    &0.49&  1.88&   18.3    &332.5& $-$21.0 &1.71&  29.0&   61.2    &30.12  &3.53   &5.35   &13.09  &0.72\\
29&     5734.9& 0.80&   2438    &0.75&  1.99&   20.4    &332.6& $-$21.6 &1.62&  40.4&   94.2    &26.92  &3.67   &5.94   &13.04  &1.76\\
30&     5784.6& 0.89&   2515    &0.69&  1.90&   20.2    &344.4& $-$10.0 &1.55&  28.4&   43.1    &29.89  &3.44   &5.27   &13.04  &1.20\\
31&     5834.0& 0.89&   2257    &0.70&  1.88&   18.8    &330.0& $-$11.3 &1.72&  38.1&   42.7    &29.92  &3.26   &4.71   &13.01  &0.52\\
32&     5982.9& 0.94&   2500    &0.65&  1.81&   20.3    &340.7& $-$17.8 &1.69&  46.3&   51.6    &31.68  &2.75   &4.65   &12.94  &0.61\\
33&     6031.9& 0.92&   2461    &0.71&  1.89&   15.6    &343.8& $-$23.0 &1.80&  37.8&   69.2    &30.91  &2.47   &5.08   &12.95  &0.55\\
34&     6032.8& 0.88&   2599    &0.75&  1.73&   16.4    &335.8& $-$26.8 &1.80&  28.5&   49.3    &29.63  &2.53   &5.77   &12.98  &0.67\\
35&     6081.8& 0.82&   2233    &0.72&  1.89&   19.8    &347.5& $-$13.3 &1.69&  49.6&   62.3    &27.41  &2.94   &4.70   &13.00  &0.48\\
36&     6131.5& 0.85&   2343    &0.75&  1.86&   14.6    &340.6& $-$28.5 &1.72&  48.1&   52.1    &28.67  &3.24   &5.37   &13.01  &0.59\\
37&     6181.5& 0.81&   2266    &0.75&  1.91&   20.5    &337.0& 28.7    &1.67&  50.5&   53.5    &27.15  &2.66   &5.40   &13.04  &0.67\\
38&     6231.5& 0.89&   2318    &0.55&  1.80&   15.4    &330.4& $-$29.9 &1.75&  45.9&   84.6    &27.84  &3.83   &5.65   &13.06  &0.68\\
39&     6379.8& 0.94&   2158    &0.64&  1.90&   18.3    &330.4& $-$13.2 &1.79&  35.9&   75.6    &31.64  &3.03   &5.81   &13.08  &0.64\\
40&     6428.8& 0.92&   2270    &0.63&  1.77&   17.1    &333.4& $-$5.8  &1.57&  33.5&   63.2    &31.00  &3.21   &5.60   &13.08  &0.54\\
41&     6454.7& 0.94&   2229    &0.64&  1.90&   19.7    &349.9& $-$29.3 &1.49&  50.7&   67.9    &31.59  &3.08   &5.47   &13.08  &0.70\\
42&     6503.6& 0.94&   2443    &0.62&  1.89&   20.4    &337.5& $-$21.7 &1.61&  41.0&   50.4    &31.63  &2.53   &5.43   &13.06  &0.83\\
43&     6726.8& 0.93&   2243    &0.71&  1.84&   20.2    &349.5& $-$16.8 &1.77&  39.1&   47.5    &31.24  &3.41   &4.58   &12.98  &0.50\\
44&     6751.9& 0.95&   2492    &0.69&  1.84&   19.4    &336.2& $-$16.0 &1.77&  47.8&   72.6    &31.85  &2.34   &4.80   &13.00  &0.55\\
45&     6776.7& 0.86&   2198    &0.69&  1.64&   18.3    &341.5& $-$28.1 &1.80&  50.1&   41.9    &28.89  &2.72   &4.72   &12.99  &0.22\\
46&     6826.1& 0.84&   2360    &0.61&  1.81&   17.6    &336.9& $-$21.4 &1.71&  41.8&   45.8    &28.18  &2.75   &4.65   &13.01  &0.42\\
47&     6900.5& 0.86&   2198    &0.69&  1.64&   18.3    &341.5& $-$28.1 &1.80&  50.1&   41.9    &28.89  &2.72   &4.72   &13.02  &0.22\\
48&     7098.8& 0.82&   2299    &0.72&  1.83&   18.8    &328.8& $-$28.1 &1.63&  48.2&   56.5    &27.61  &3.33   &5.67   &13.07  &0.76\\
49&     7098.9& 0.78&   2459    &0.54&  1.90&   17.8    &323.6& $-$18.9 &1.70&  27.4&   43.0    &26.35  &3.47   &5.31   &13.10  &1.00\\
50&     7173.7& 0.86&   2486    &0.74&  1.87&   18.6    &335.3& $-$29.5 &1.74&  27.2&   58.7    &28.93  &3.09   &6.40   &13.06  &1.32\\
51&     7197.9& 0.89&   2362    &0.74&  1.90&   17.1    &335.4& $-$29.9 &1.75&  32.0&   58.4    &29.82  &3.45   &5.89   &13.04  &0.99\\
52&     7222.8& 0.95&   2444    &0.63&  1.92&   17.5    &338.3& $-$28.9 &1.76&  26.5&   51.0    &31.98  &2.99   &5.43   &13.04  &0.89\\
\hline
Mean & &0.91&   2543 &0.53&1.83&18.6&335.6&$-$16.3&1.65&40.8&59.1&30.70&3.00&5.34&13.03&1.02\\
std &  &0.05 &267 &0.17&0.10&1.6&8.4&14.0&0.13&8.0&15.7&1.80 &0.78&0.67&0.05&0.90\\
\hline
\vspace{0.05cm}
\end{tabularx}
Note: The meaning of the parameters is explained in the text. The mass transfer rate is normalized to the value of LC= 49, where $R_d$ attains its minimum value. Typical errors are given along with the means and standard deviations.
\end{table*}

\begin{figure*}
\scalebox{1}[1]{\includegraphics[angle=0,width=17cm]{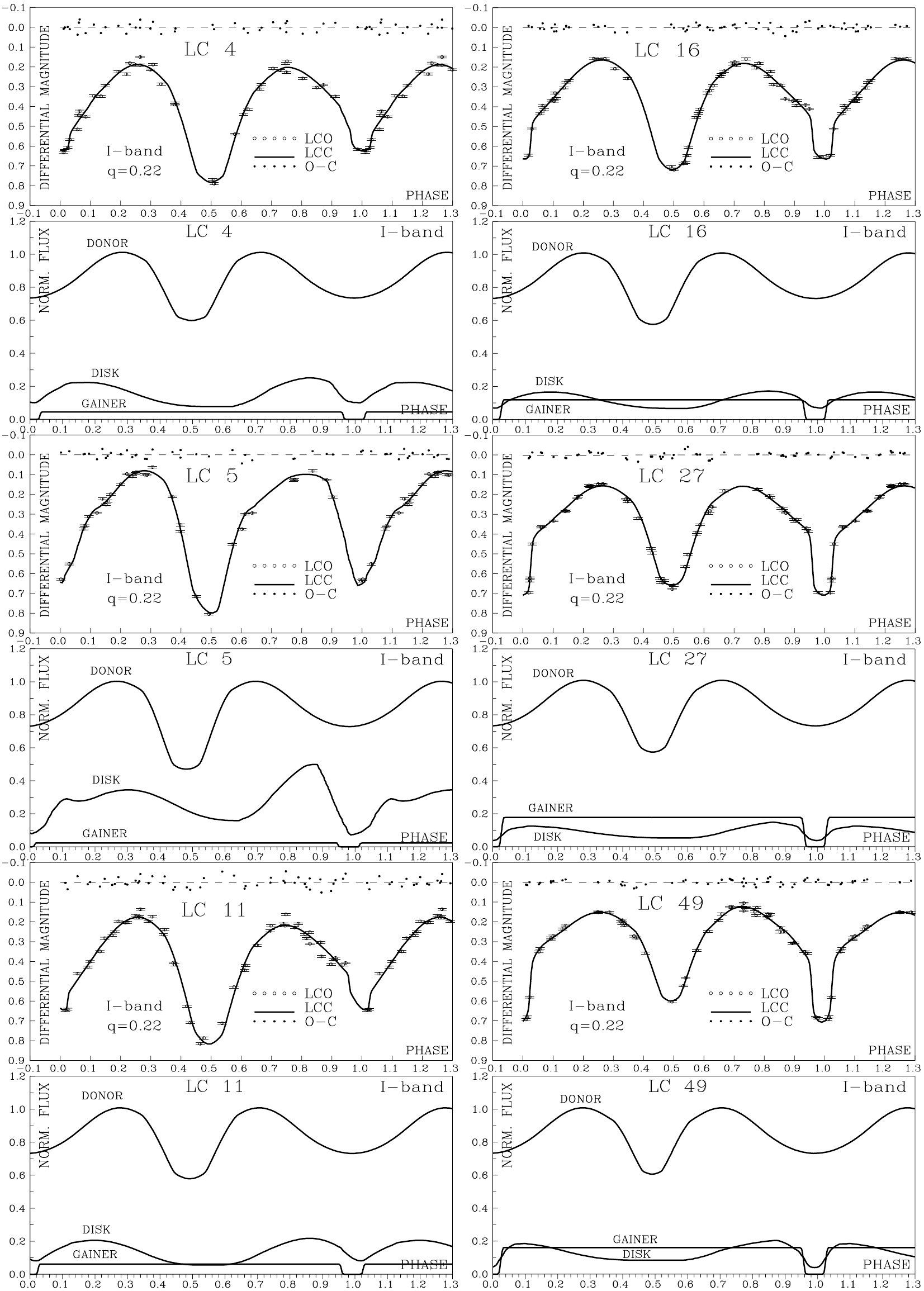}}
\caption{ Some light curves showing large mass transfer rates (LC = 4, 5, 11), along with others used for comparison (LC = 16, 27, 49). Relative fluxes of donor, disk, and gainer are shown versus the orbital phase.}
\label{fig:fignew}
\end{figure*}

\end{appendix}

\end{document}